\input{aipcheck}

\documentclass[
    ,final            % use final for the camera ready runs
  ]
  {aipproc}

\layoutstyle{8x11double}

\begin{document}

\title{Yang-Mills streamlines and semi-classical confinement}


\author{Kurt Langfeld}{
  address={School of Computing \& Mathematics, University of Plymouth, Plymouth Pl4 8AA, UK  }
}

\author{Ernst-Michael Ilgenfritz}{
  address={Institut f\"ur Physik, Humboldt-Universit\"at zu Berlin, D-12489 Berlin, Germany }
}

%\author{<author3>}{
%  address={<common address for author2 and author3>}
%  ,altaddress={<author1 address>} % additional visiting address
%}

\begin{abstract}
Semi-classical configurations in Yang-Mills theory have been derived from
lattice Monte Carlo configurations using a recently proposed constrained
cooling technique which is designed to preserve every Polyakov line
(at any point in space-time in any direction). Consequently, confinement
was found sustained by the ensemble of semi-classical configurations.
The existence of gluonic and fermionic near-to-zero modes was demonstrated
as a precondition for a possible semi-classical expansion around the cooled
configurations as well as providing the gapless spectrum of the Dirac
operator necessary for chiral symmetry breaking.
The cluster structure of topological charge of the semi-classical streamline
configurations was analysed and shown to support the axial anomaly of the
right size, although the structure differs from the instanton gas or liquid.
Here, we present further details on the space-time structure and the time
evolution of the streamline configurations.
\end{abstract}

\maketitle

%%%%%%%%%%%%%%%%%%%%%%%%%%%%%%%%%%%%%%%%%%%%
%% MAINMATTER
%%%%%%%%%%%%%%%%%%%%%%%%%%%%%%%%%%%%%%%%%%%%

%\section{Introduction}

\textbf{Introduction:}
The lattice approach has been very successful to detail the
low energy properties of QCD as encoded in the hadron spectrum
and hadron structure. The success of the vortex and monopole
picture~\cite{Greensite:2003bk} feeds the hope that many features of the
vacuum structure, such as quark (de-) confinement,
spontaneous chiral symmetry breaking (and restoration) and the
(explicit) axial anomaly, can be understood as coming from a common
%% EMI source -> Origin  KL: source ist besser (im sinne von Herkunft)
source.
%% EMI Such a
A qualitative explanation of these phenomenawould be helpful for
%% KL added: the
the understanding of the phase structure of QCD under extreme conditions where
direct simulations are prevented by the notorious sign-problem.
%%
%% KL: raus wegen Platzmanegl
%% This understanding of the mechanism and its modification at high density
%% might also help to explain the behaviour of matter in a dense and deconfined
%%  state  when our present understanding of the QCD phase digram is
%%  challenged by heavy ion collision experiments.

\vskip 0.1cm
In the late seventies the hope was stirred that the instanton
solutions~\cite{Belavin:1975fg} could be {\it the} building blocks to
construct a semi-classical model to explain all these outstanding
features~\cite{Callan:1977gz,Callan:1978bm}. One of the aims was to find
an effective
description in terms of an enormously reduced number of degrees of
freedom and a appropriate probability measure for
them~\cite{Schafer:1996wv}.
A characteristic aspect of this model was that, in absence of quantum
fluctuations, action would be nearly and topological charge would be
strictly quantized. The original hope was not fulfilled as far as the
model failed to explain confinement.

\vskip 0.1cm
Analyzing lattice fields,
one has understood in the mean time (with filtering techniques based on
certain scalar and fermionic
operators~\cite{Bruckmann:2009vb,Bruckmann:2009qz}) that at a certain momentum
scale of resolution the topological density seems to exist in the form of
instanton-like structures. A careful tuning of cooling or smearing applied
to the gauge field can reproduce this structure. Unfortunately, it turned out,
using suitably stopped standard cooling, that the reconstructed
{\it apparent} instanton configurations can explain the spontaneous breakdown
of chiral  symmetry but likewise fail to confine
quarks~\cite{DeGrand:1997sd,Kovacs:1999an}.

\vskip 0.1cm
The most popular picture both of confinement and chiral symmetry breaking
rests on vortices (for a review see~\cite{Greensite:2003bk})
defined by the maximal centre gauge~\cite{DelDebbio:1996mh}.
Although networks of thin, percolating centre vortices are infinite action
configurations, their properties nicely scale with the lattice regulator
granting the vortex configurations a meaningful interpretation in the
continuum limit~\cite{Langfeld:1997jx}. It was pointed out
in~\cite{Gubarev:2002ek,Zakharov:2003vh} that the balance between
vortex entropy and vortex action must be intrinsically fine-tuned
in Yang-Mills theory to explain these scaling properties.
A natural explanation for this fine-tuning was firstly offered by one
of us in~\cite{Langfeld:2009es}: the vortices are the singular image,
in a certain gauge, of otherwise smooth semi-classical configurations
which confine and which, this time, are probably not built out of
instantons. In our recent paper~\cite{Langfeld:2010nm}, we
have expanded the cooling techniques and did find ensembles of
semi-classical configurations which confine quarks.
We here summarise the ideas and main results of this approach and
present new result on the space-time structure of these configurations.

\vskip 0.1cm
\textbf{ Constrained cooling:}
Recently, we supplemented cooling of lattice configurations
with certain constraints expressing conditions valid in the
infrared~\cite{Langfeld:2009es,Langfeld:2010nm}.
This should allow us to let cooling run non-supervised, providing
ensembles of genuinely
semi-classical configurations which conserve the property of confinement.
%% EMI added "Originally"  Frage: ist "referred" richtig ? KL: Ja
Originally, the name ``streamline'' referred to instanton anti-instanton
pairs living in concurrence with the empty perturbative
vacuum~\cite{Verbaarschot:1991sq}. These pairs would annihilate under cooling,
if the constraints were not in place.

\vskip 0.1cm
The idea in~\cite{Langfeld:2010nm} was to reduce the action of $SU(2)$
lattice configurations with a huge number of constraints to preserve
each Polyakov loop on the lattice. It might be 

\begin{figure}[h!]
\begin{tabular}{cccc}
  \includegraphics[width=.18\textwidth]{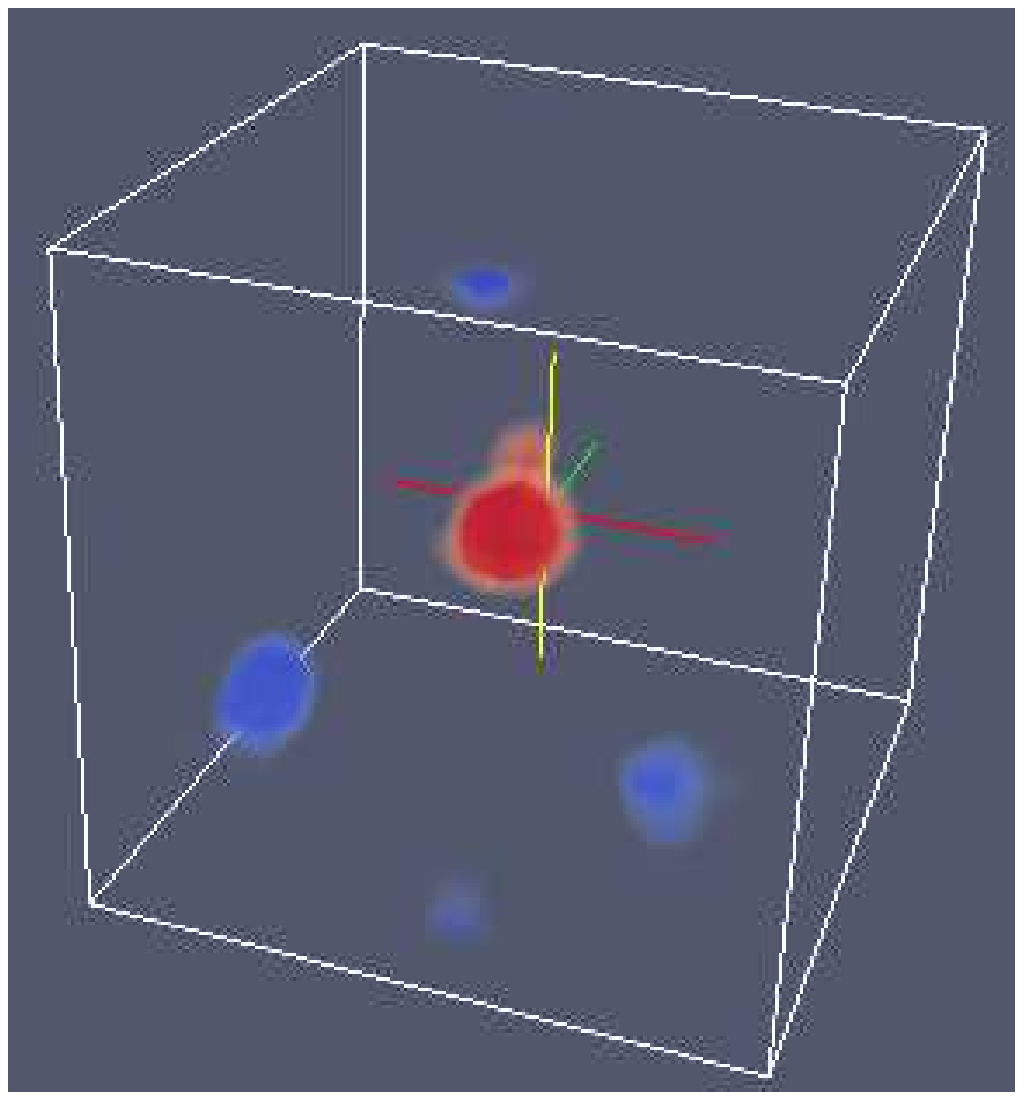}  &
  \includegraphics[width=.18\textwidth]{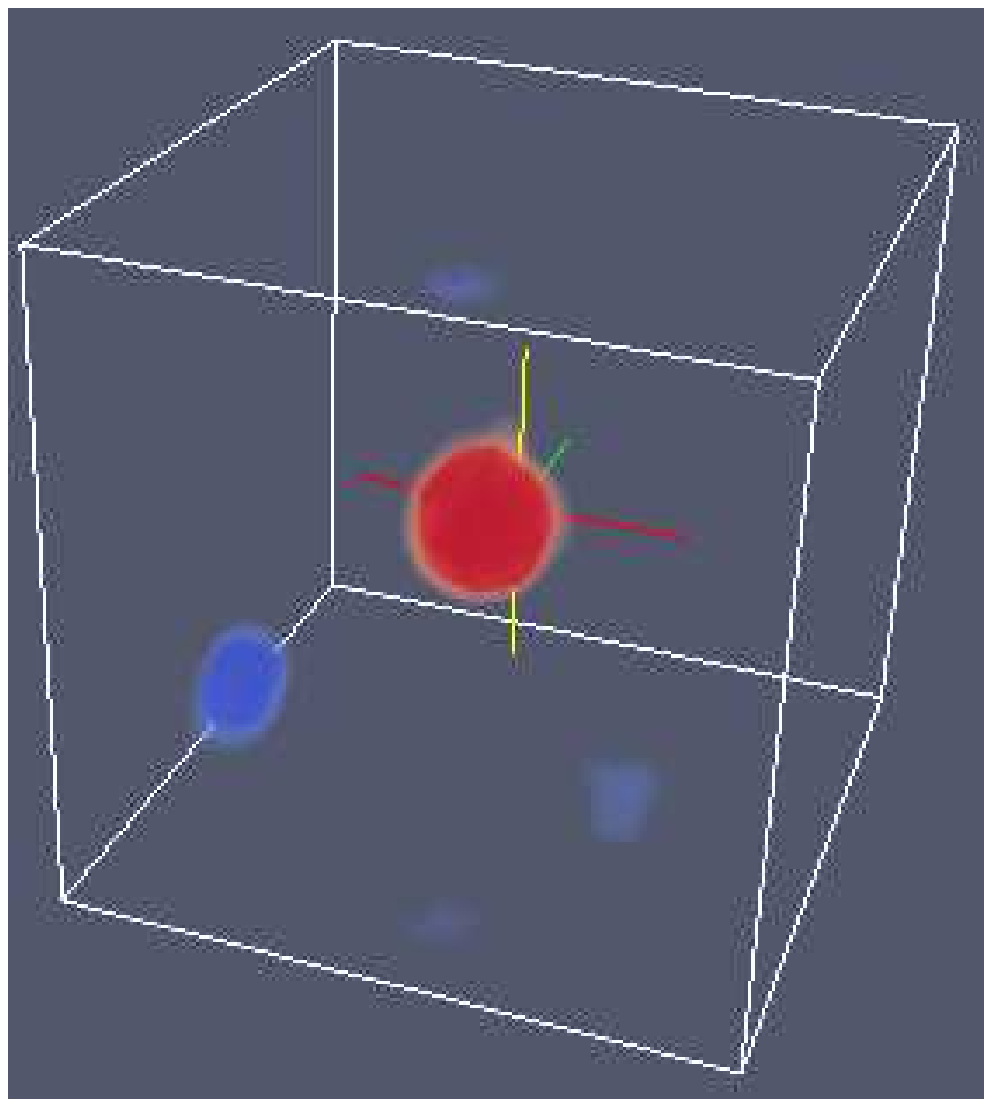}  &
  \includegraphics[width=.18\textwidth]{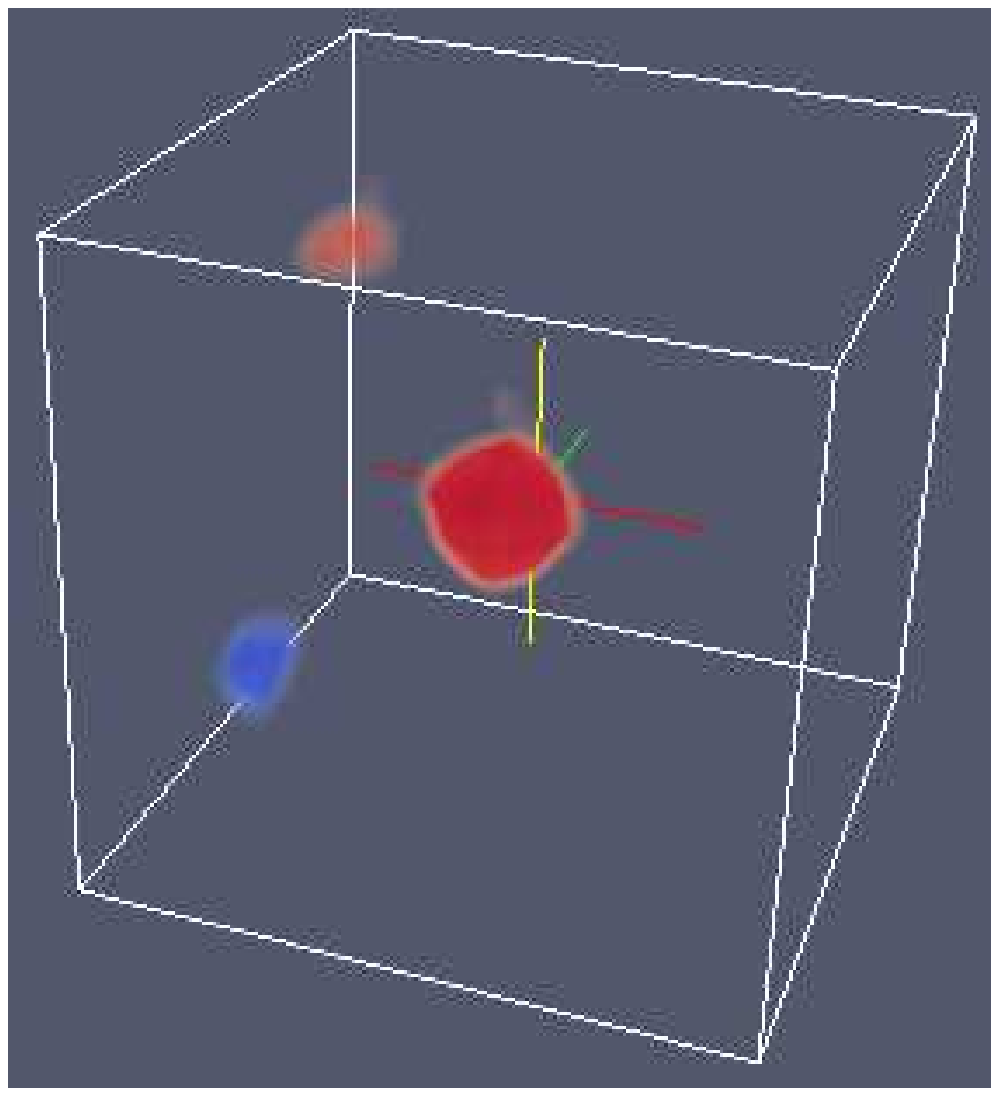}  &
  \includegraphics[width=.18\textwidth]{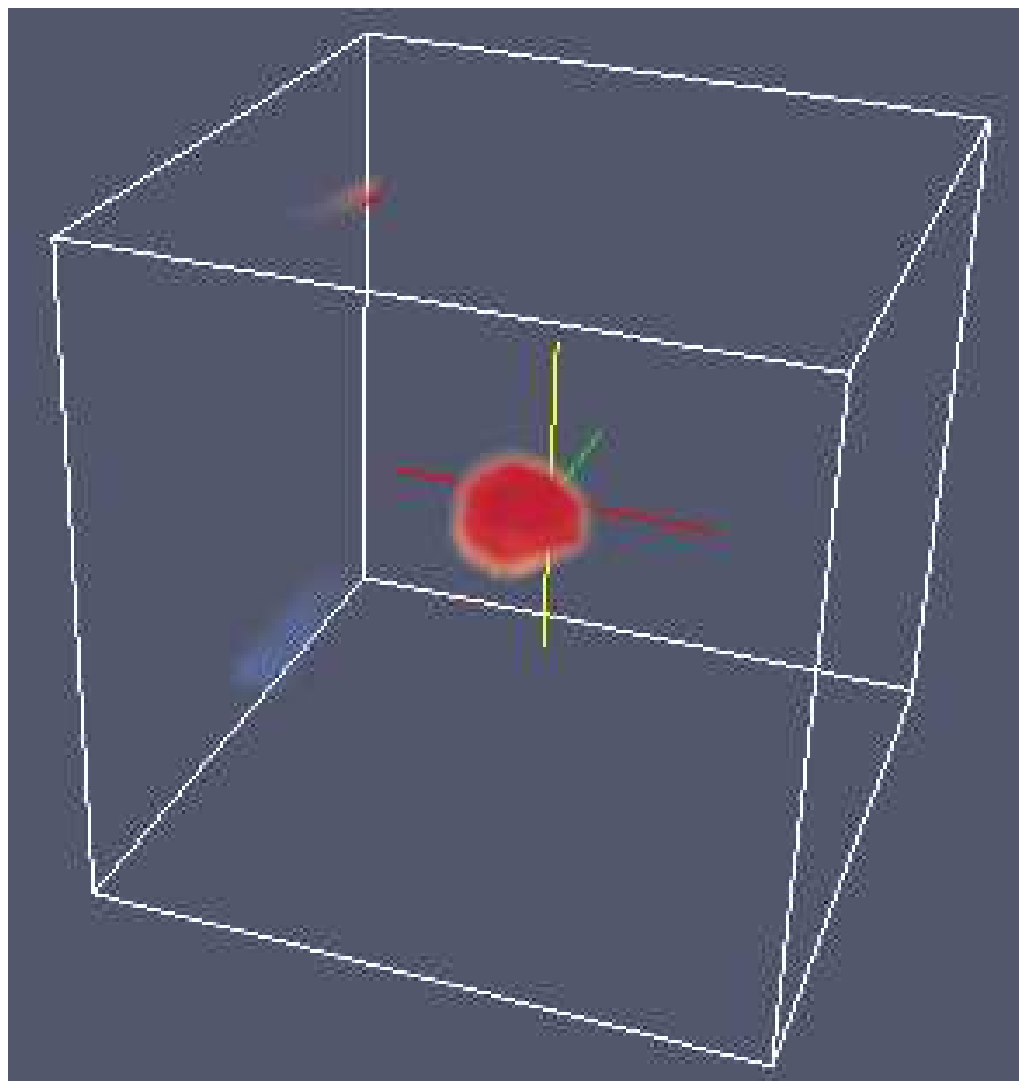}  \\
  \includegraphics[width=.18\textwidth]{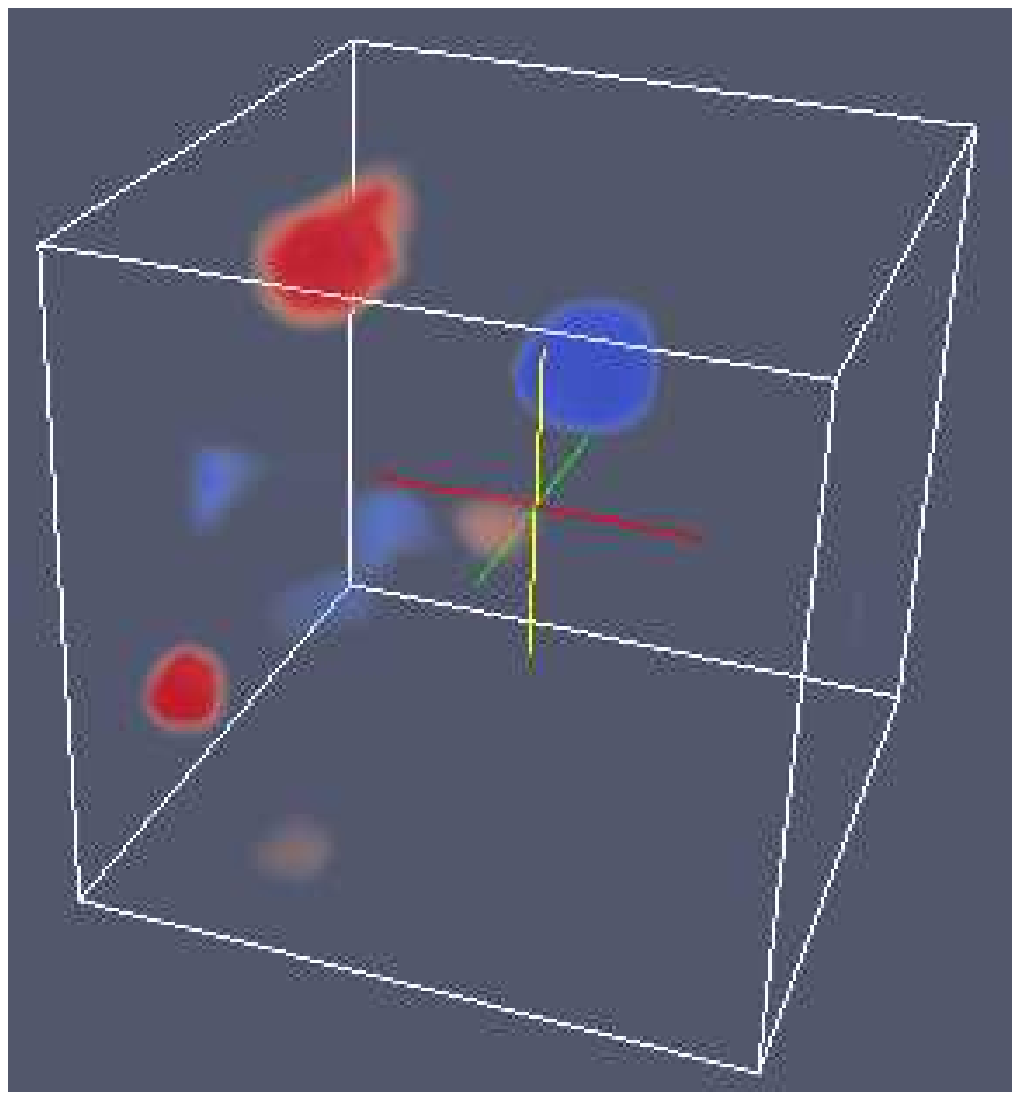}  &
  \includegraphics[width=.18\textwidth]{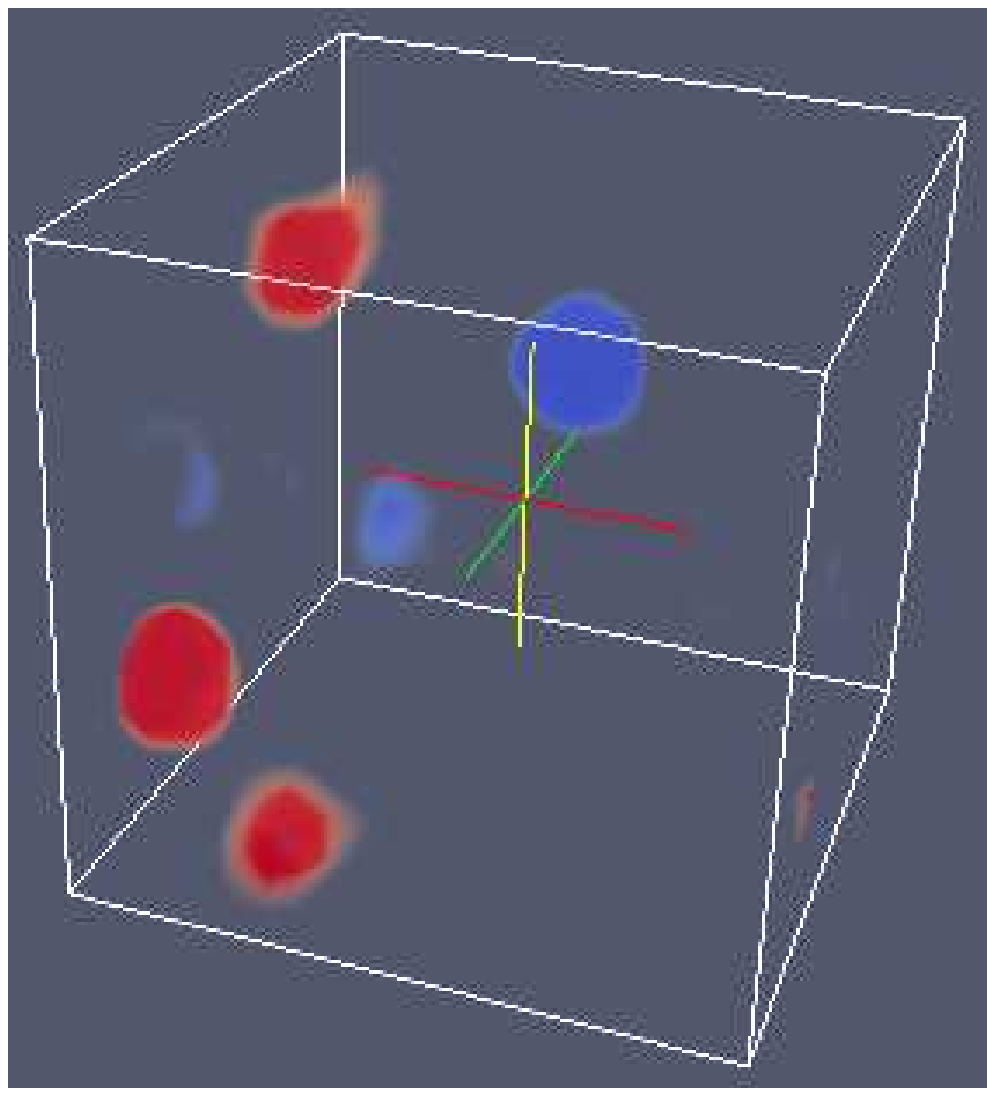}  &
  \includegraphics[width=.18\textwidth]{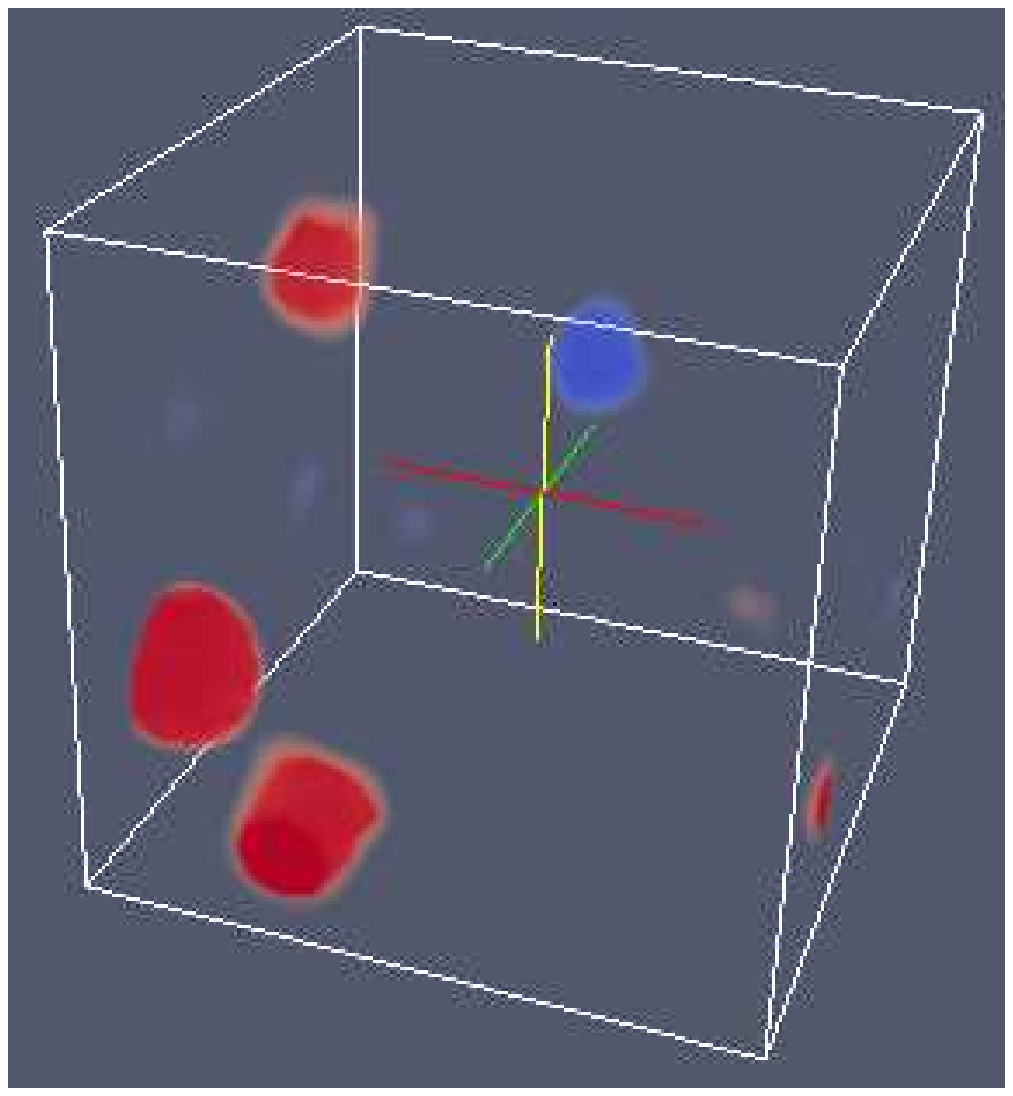}  &
  \includegraphics[width=.18\textwidth]{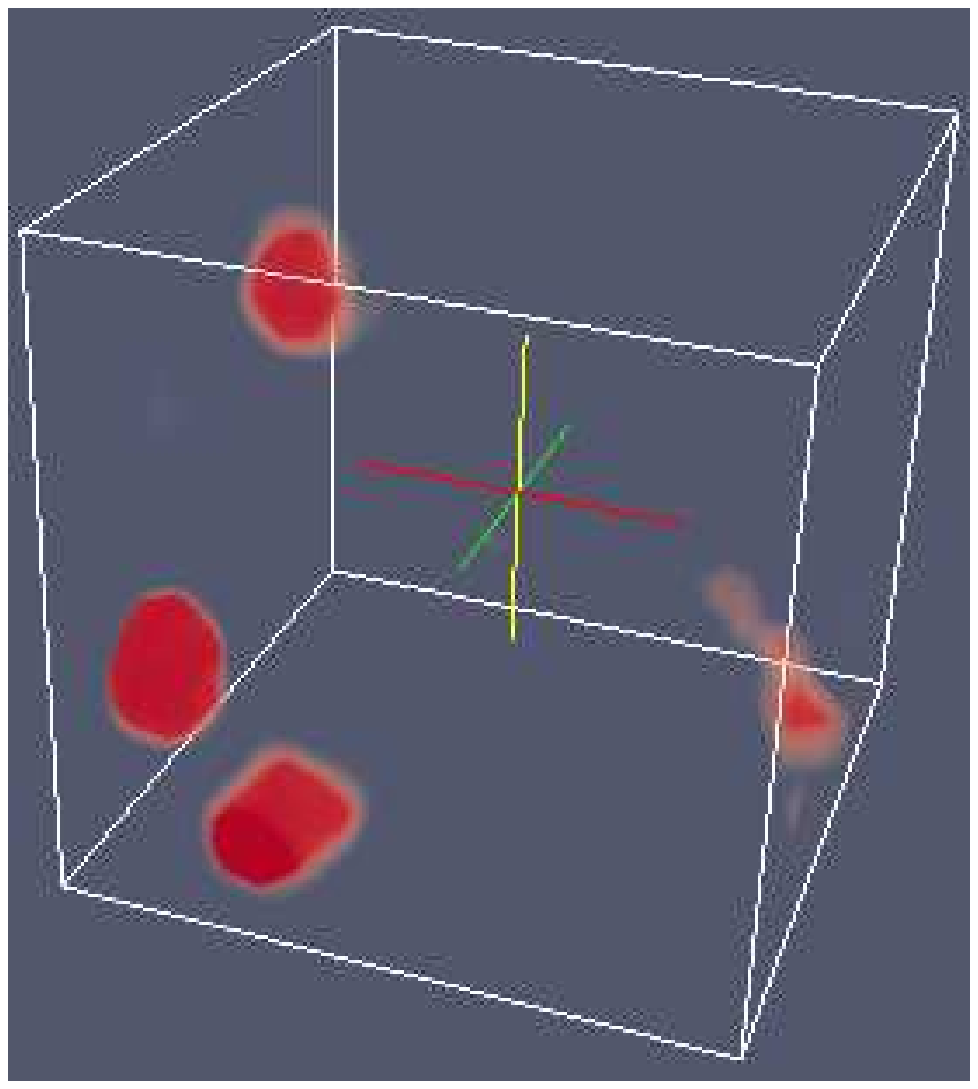}  \\
  \includegraphics[width=.18\textwidth]{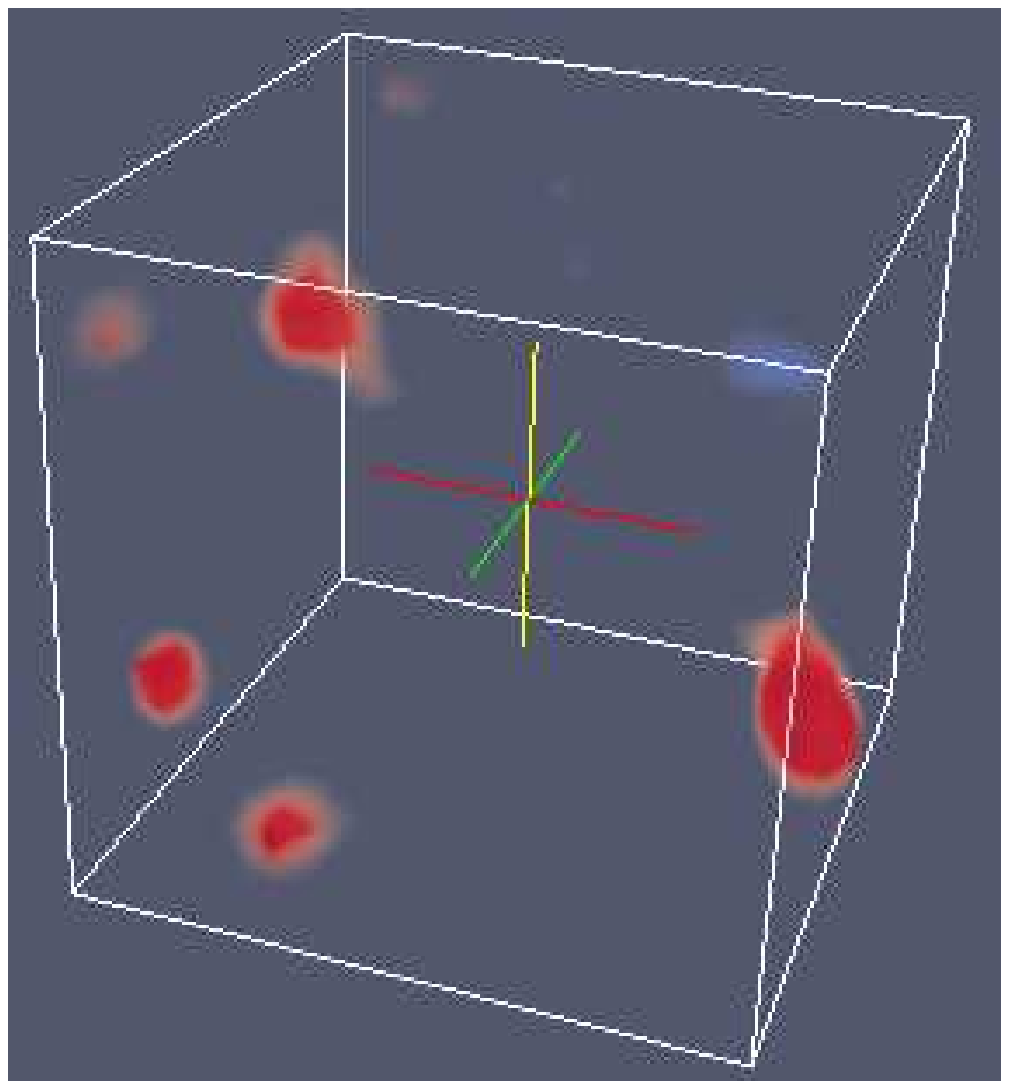}  &
  \includegraphics[width=.18\textwidth]{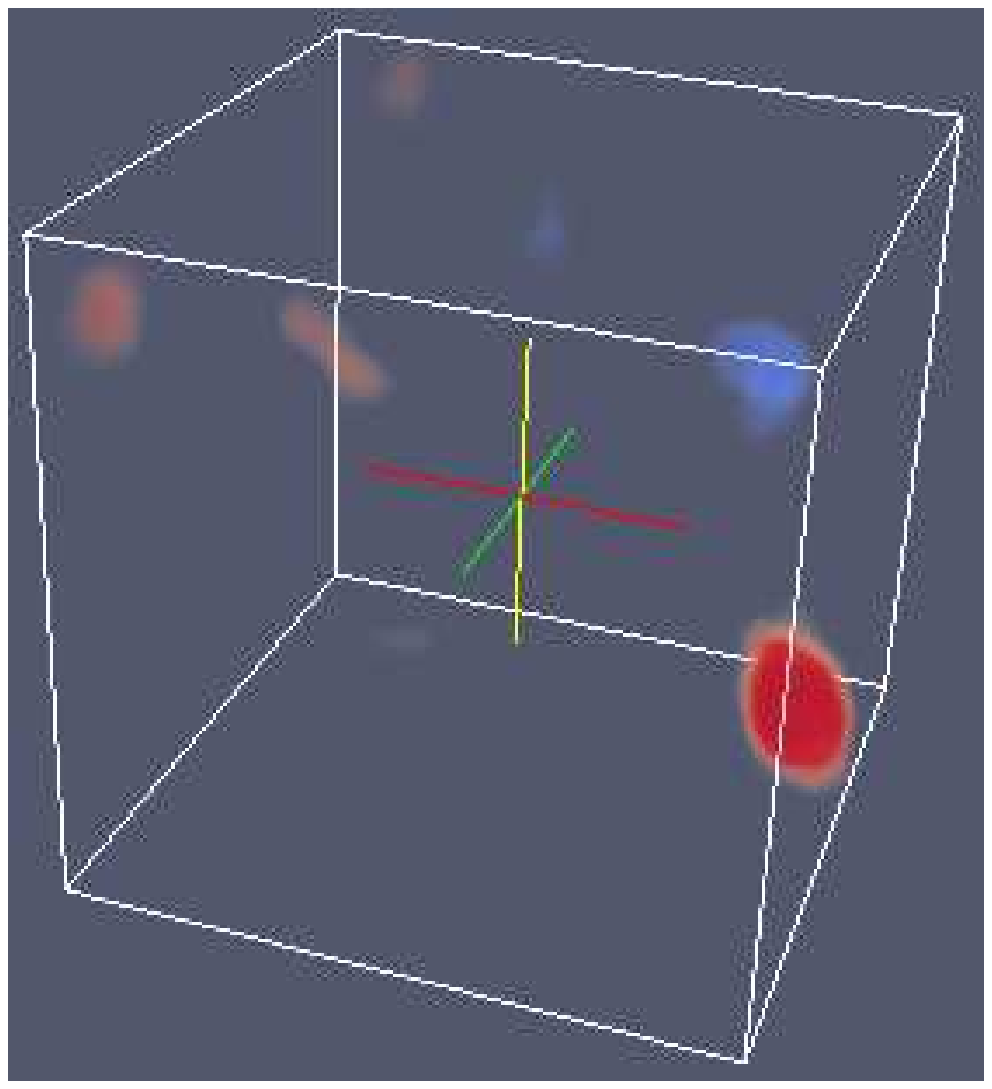}  &
  \includegraphics[width=.18\textwidth]{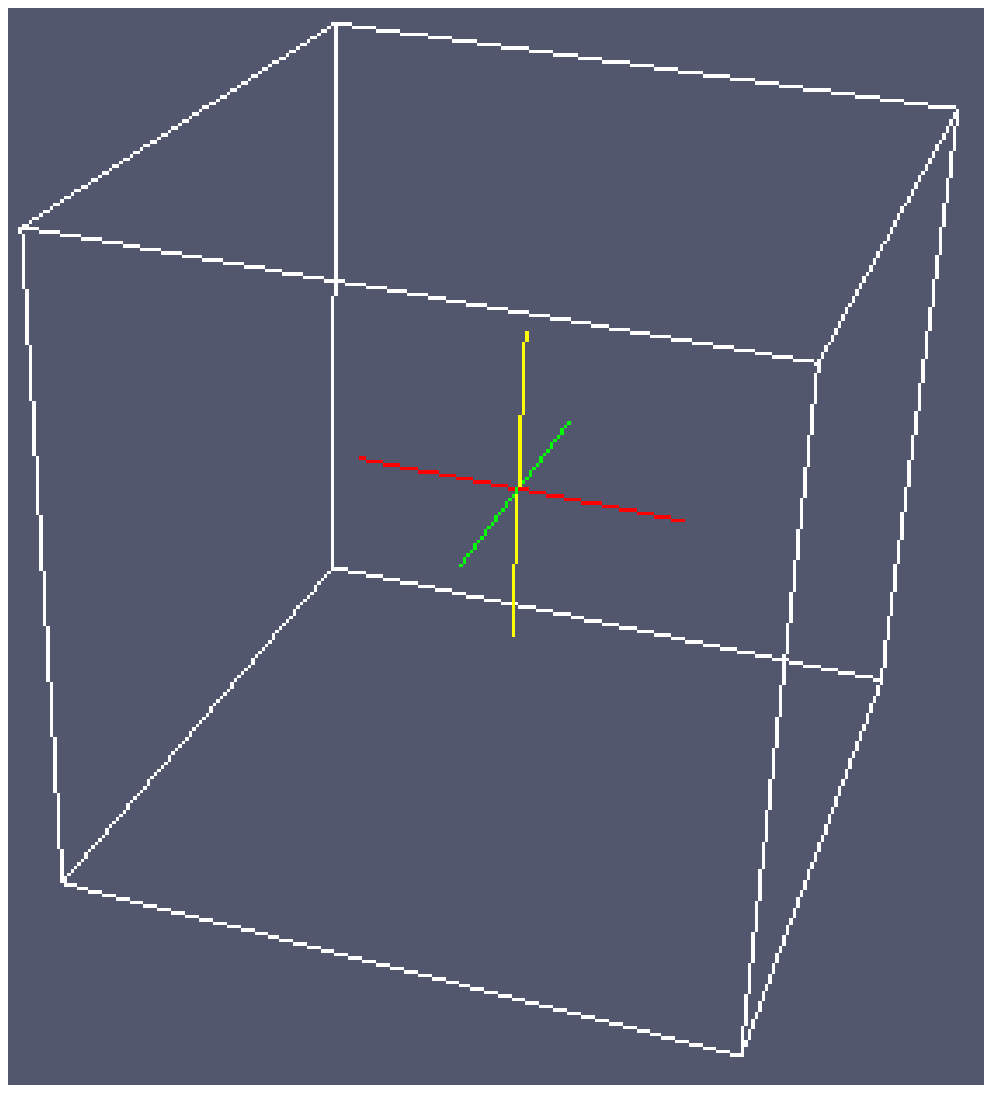}  &
  \includegraphics[width=.18\textwidth]{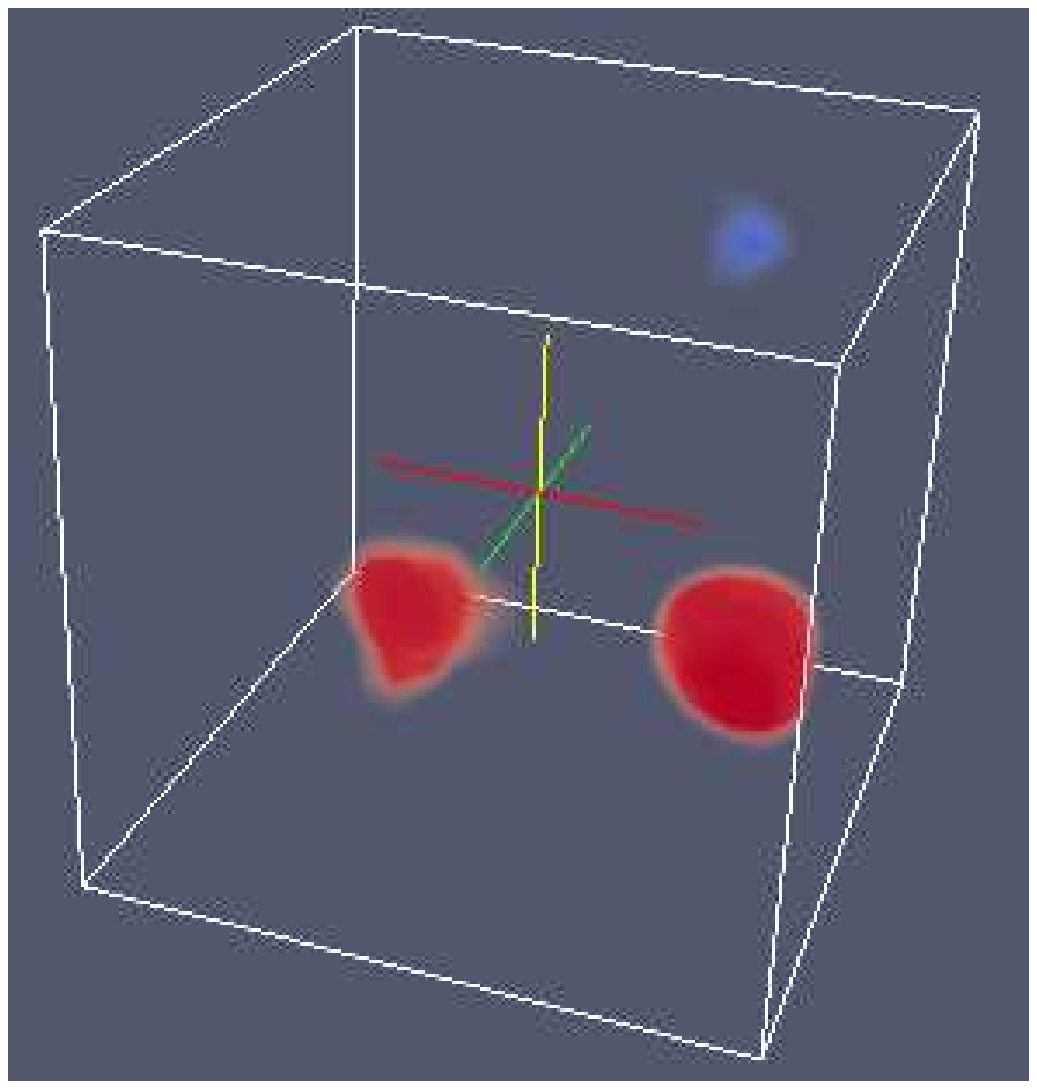}  \\
  \includegraphics[width=.18\textwidth]{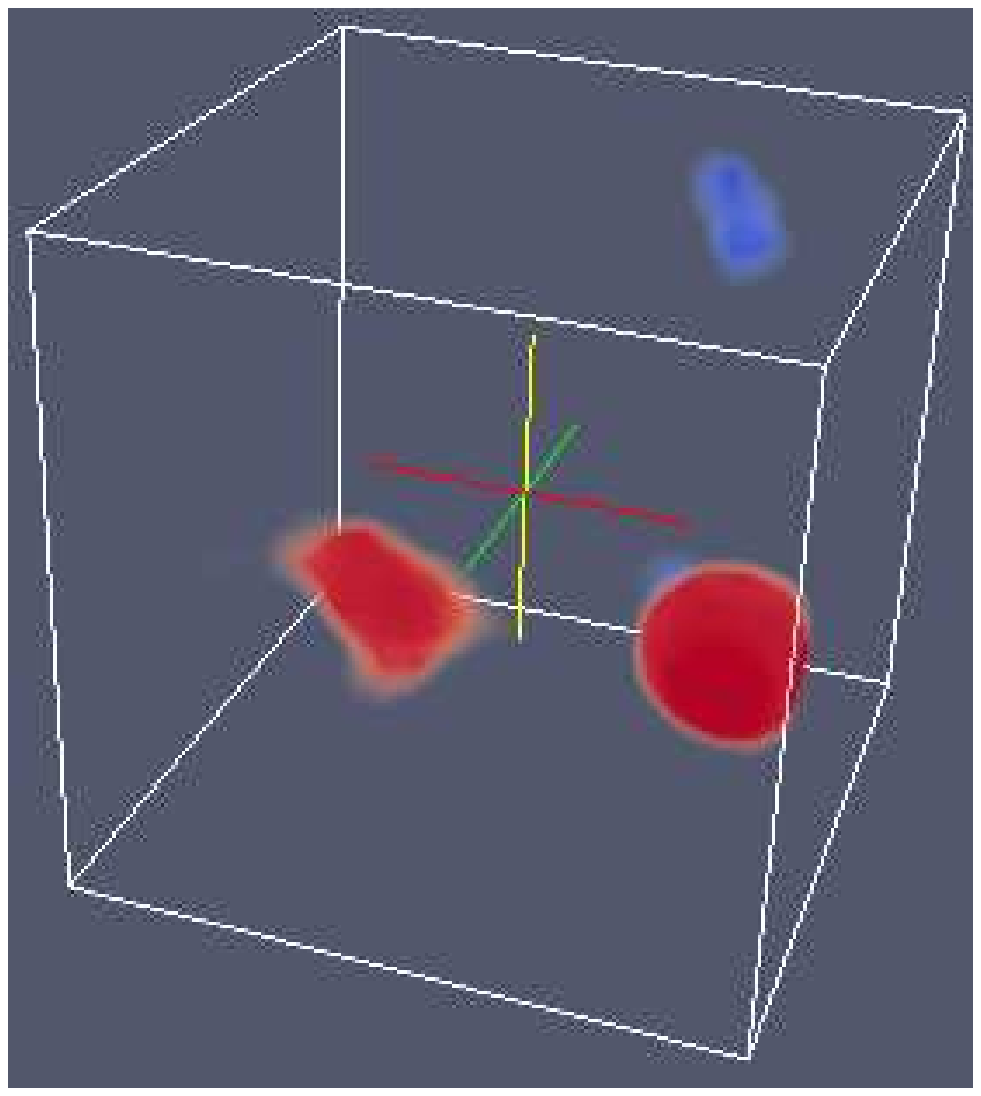}  &
  \includegraphics[width=.18\textwidth]{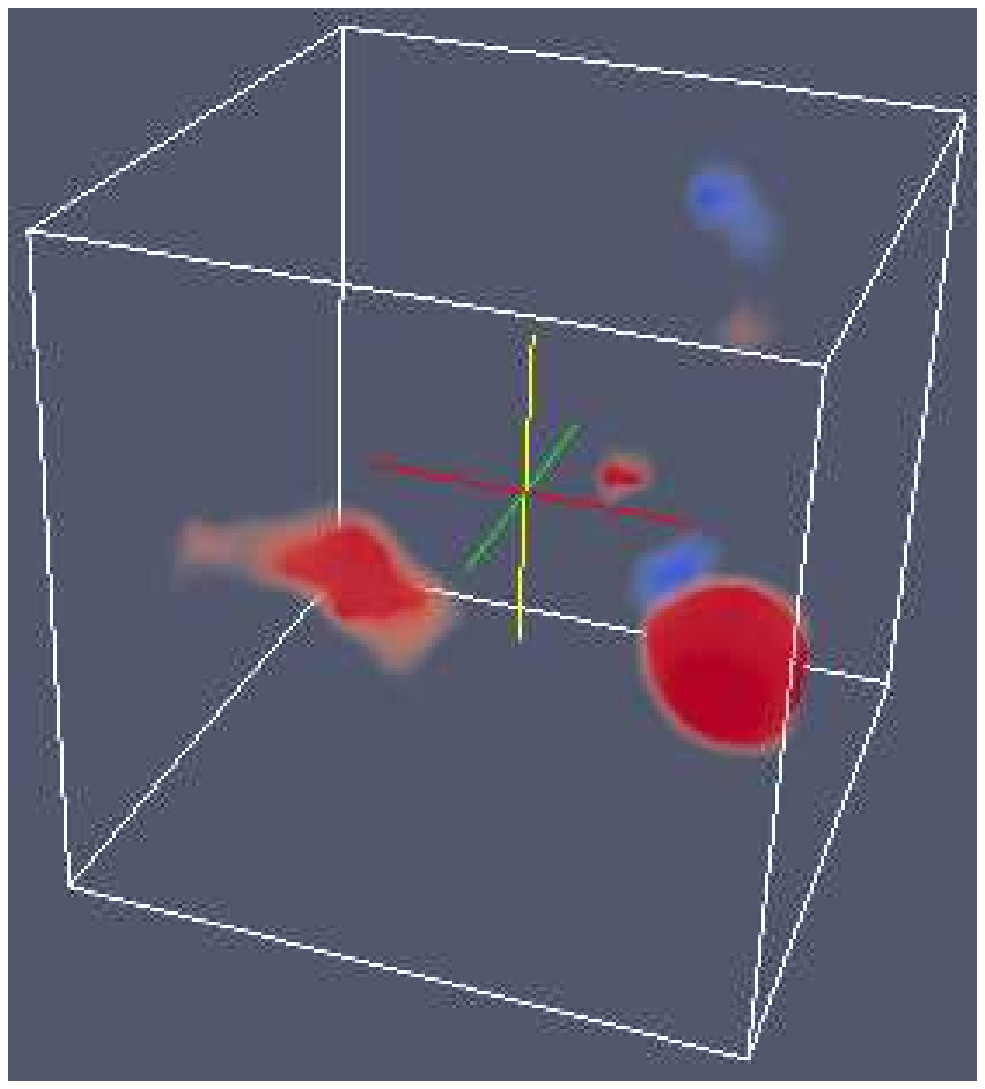}  &
  \includegraphics[width=.18\textwidth]{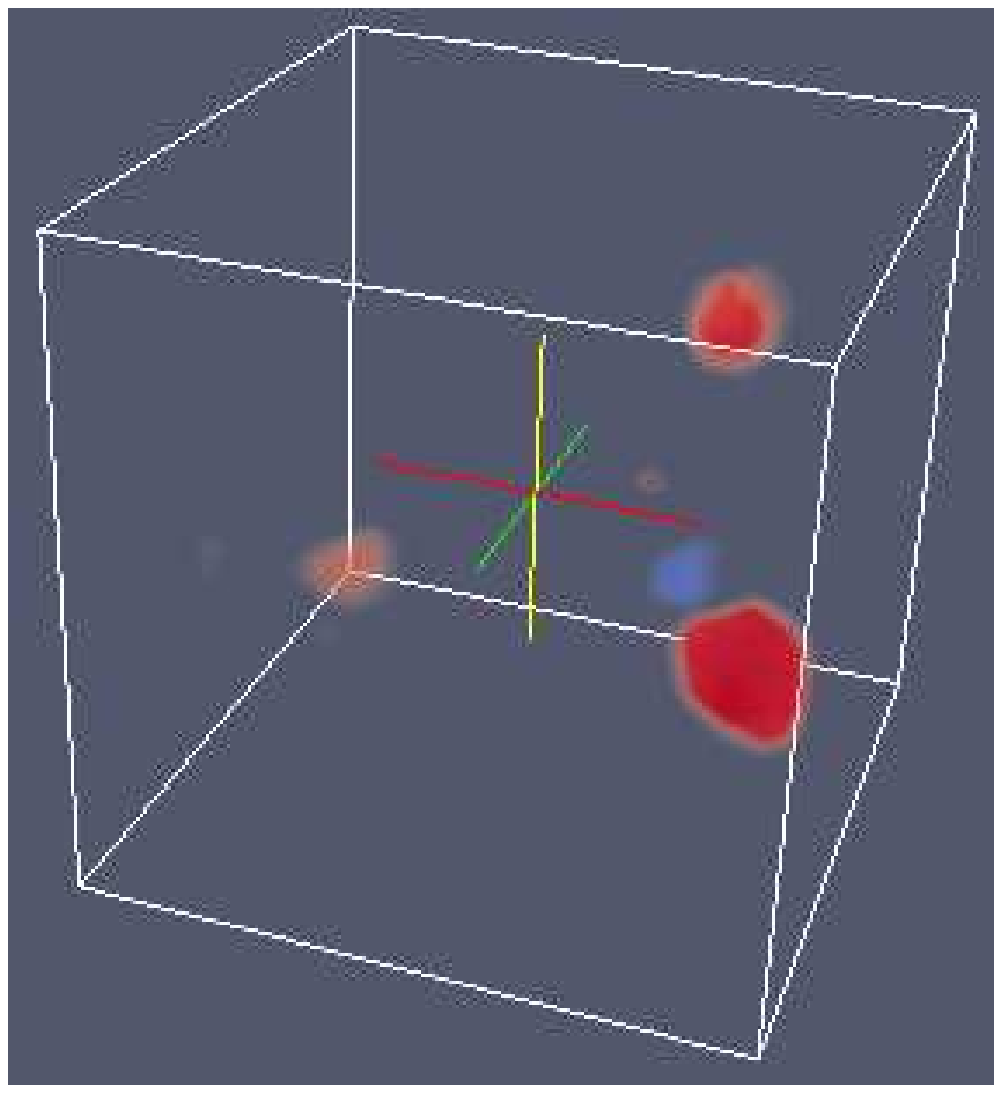}  &
  \includegraphics[width=.18\textwidth]{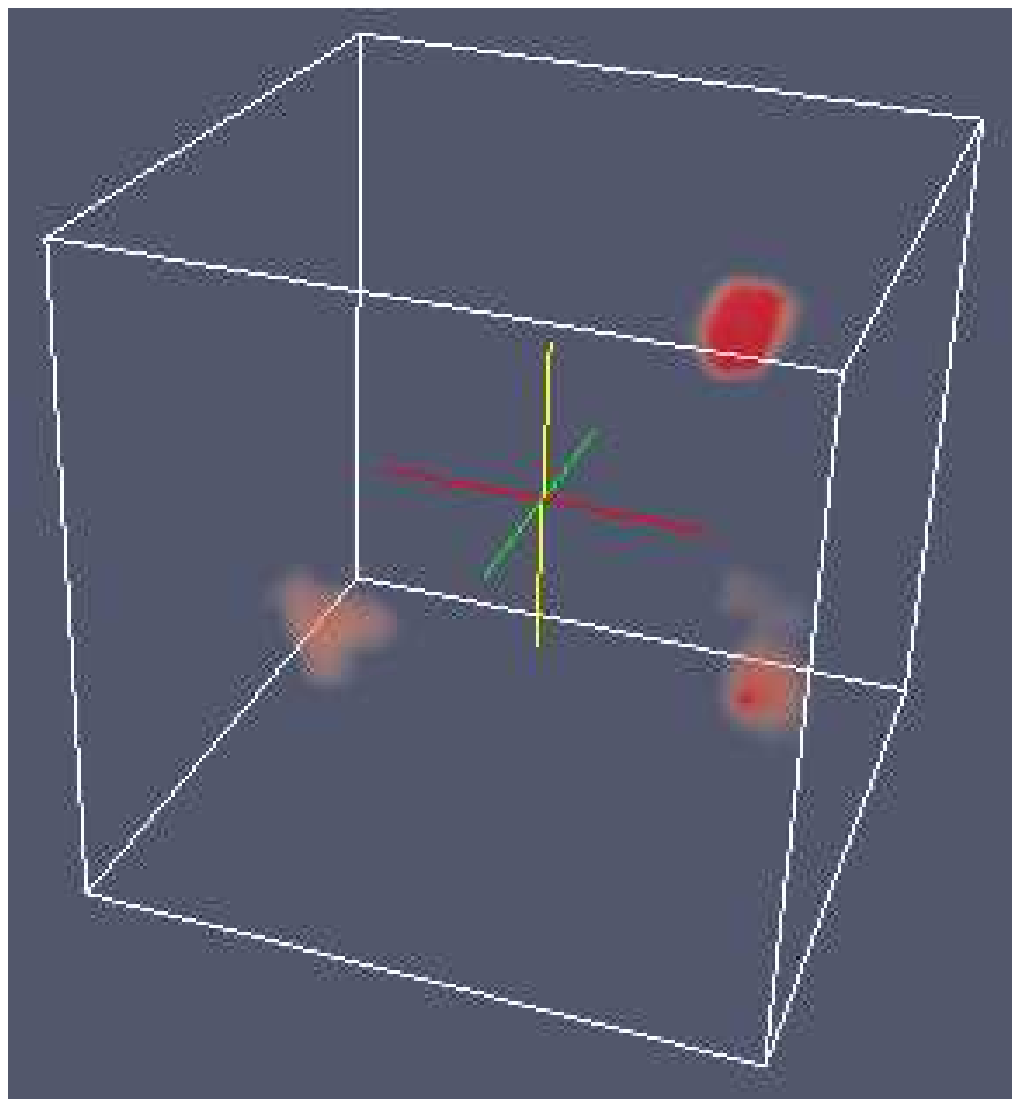}  \\
  \includegraphics[width=.18\textwidth]{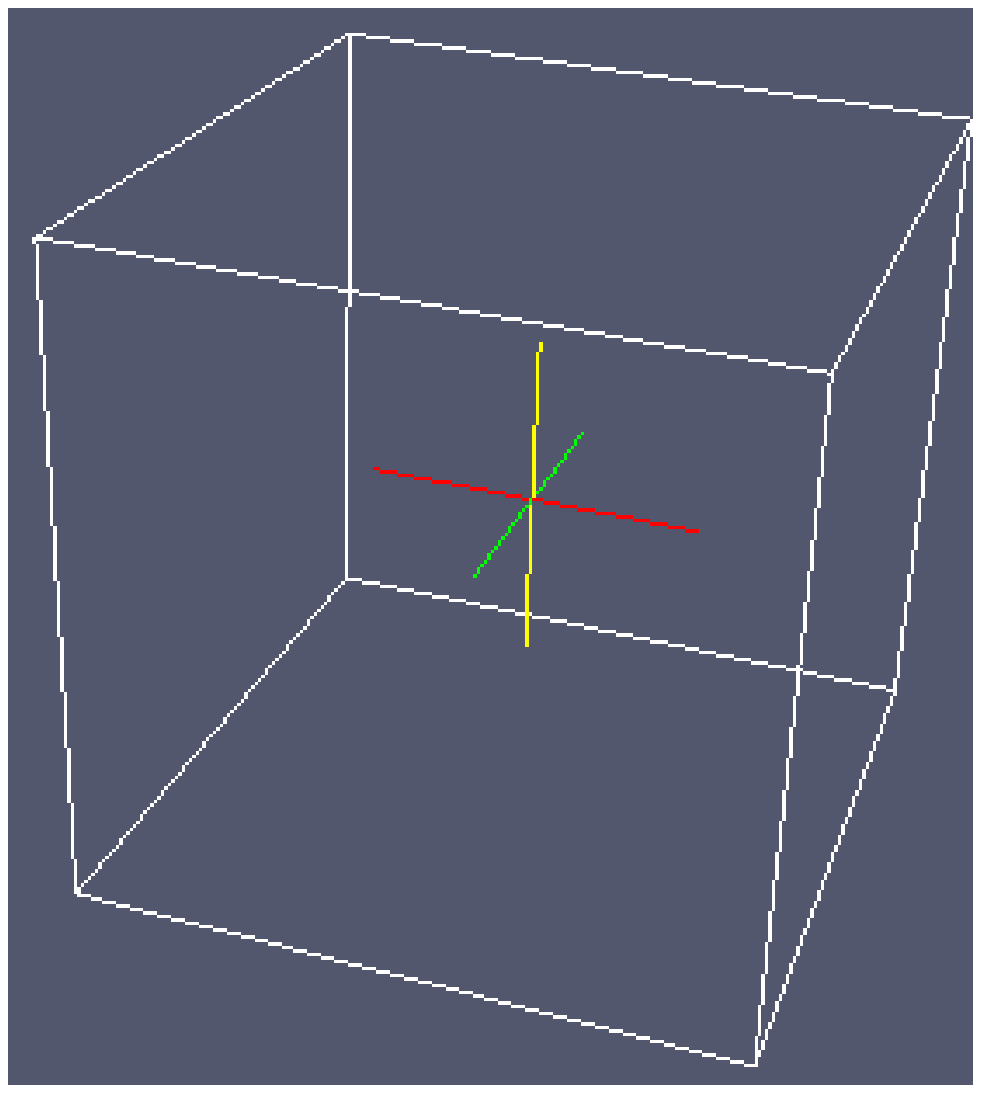}  &
  \includegraphics[width=.18\textwidth]{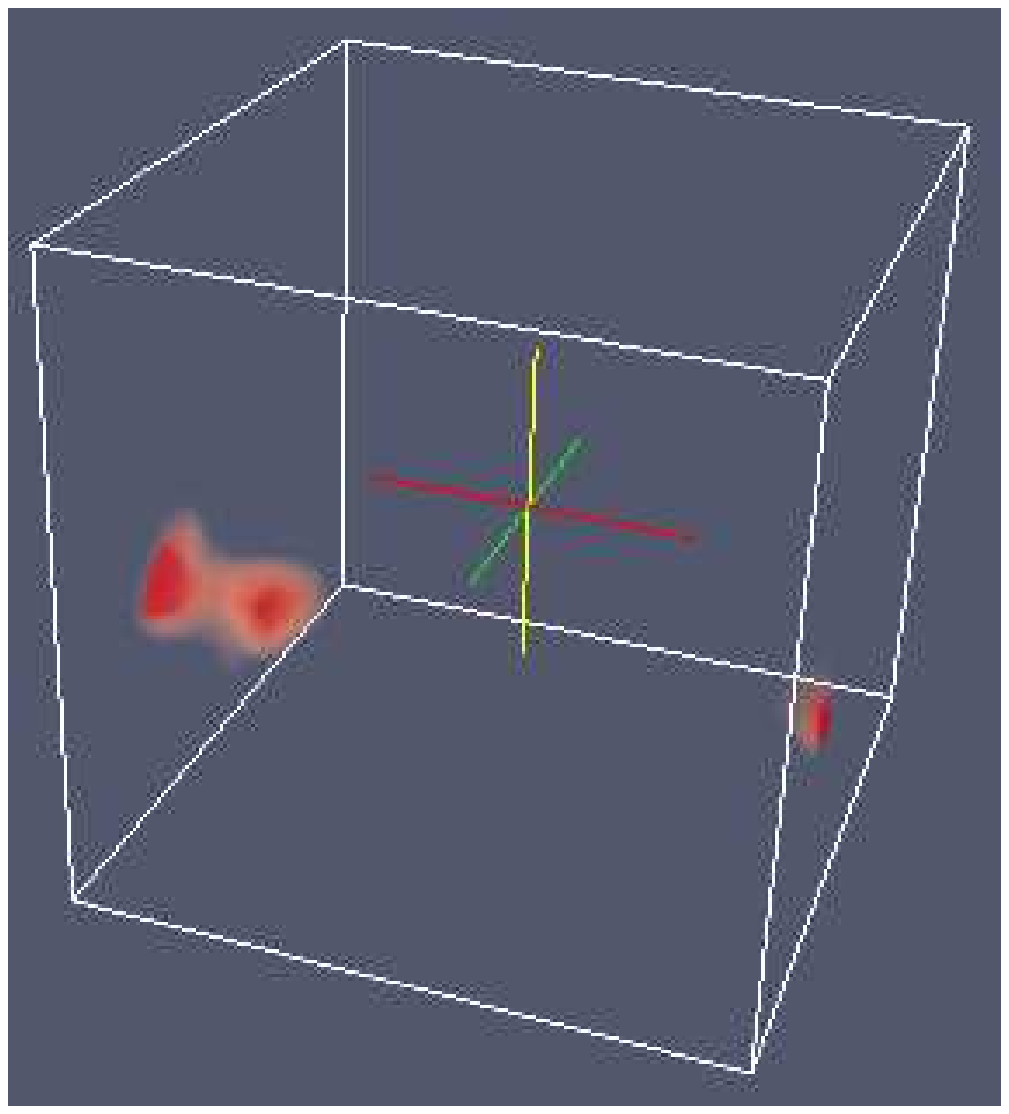}  &
  \includegraphics[width=.18\textwidth]{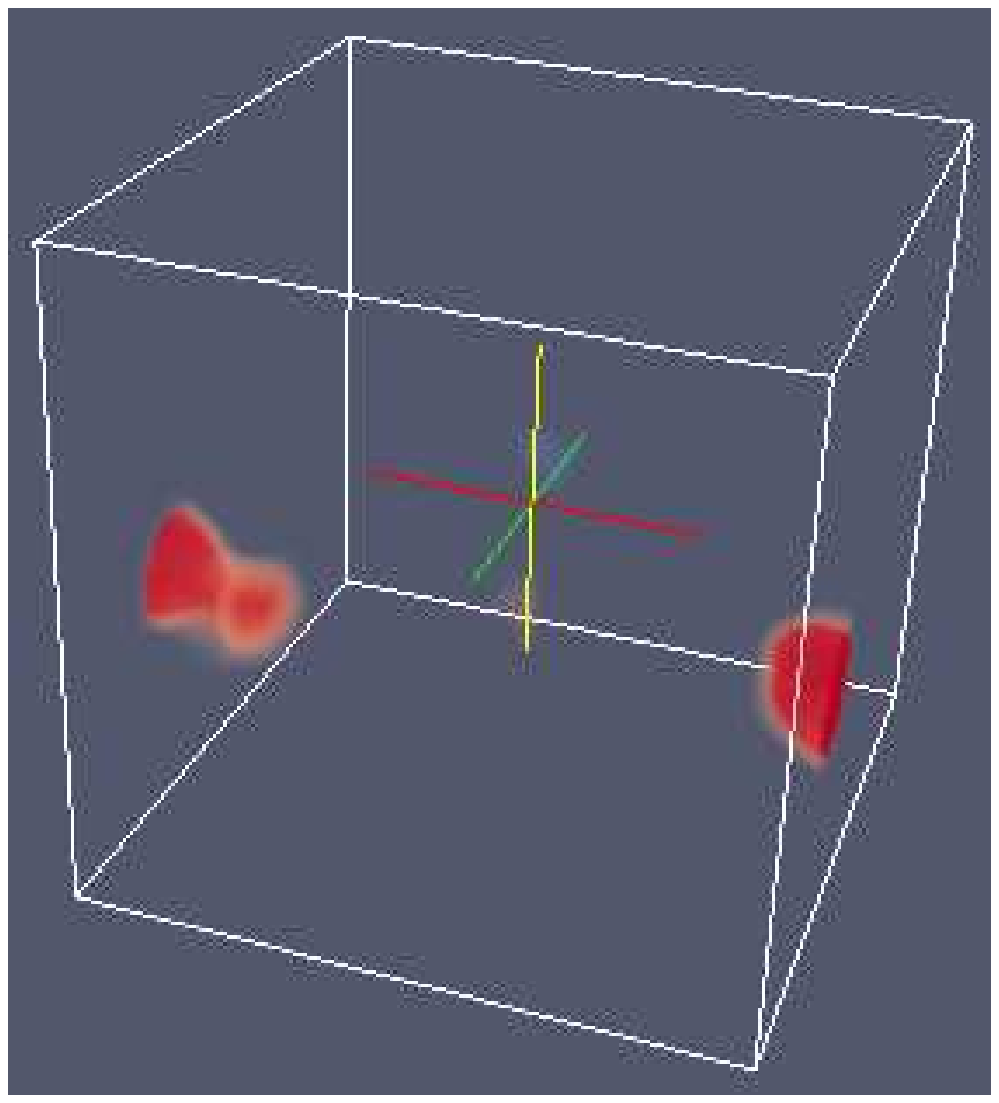}  &
  \includegraphics[width=.18\textwidth]{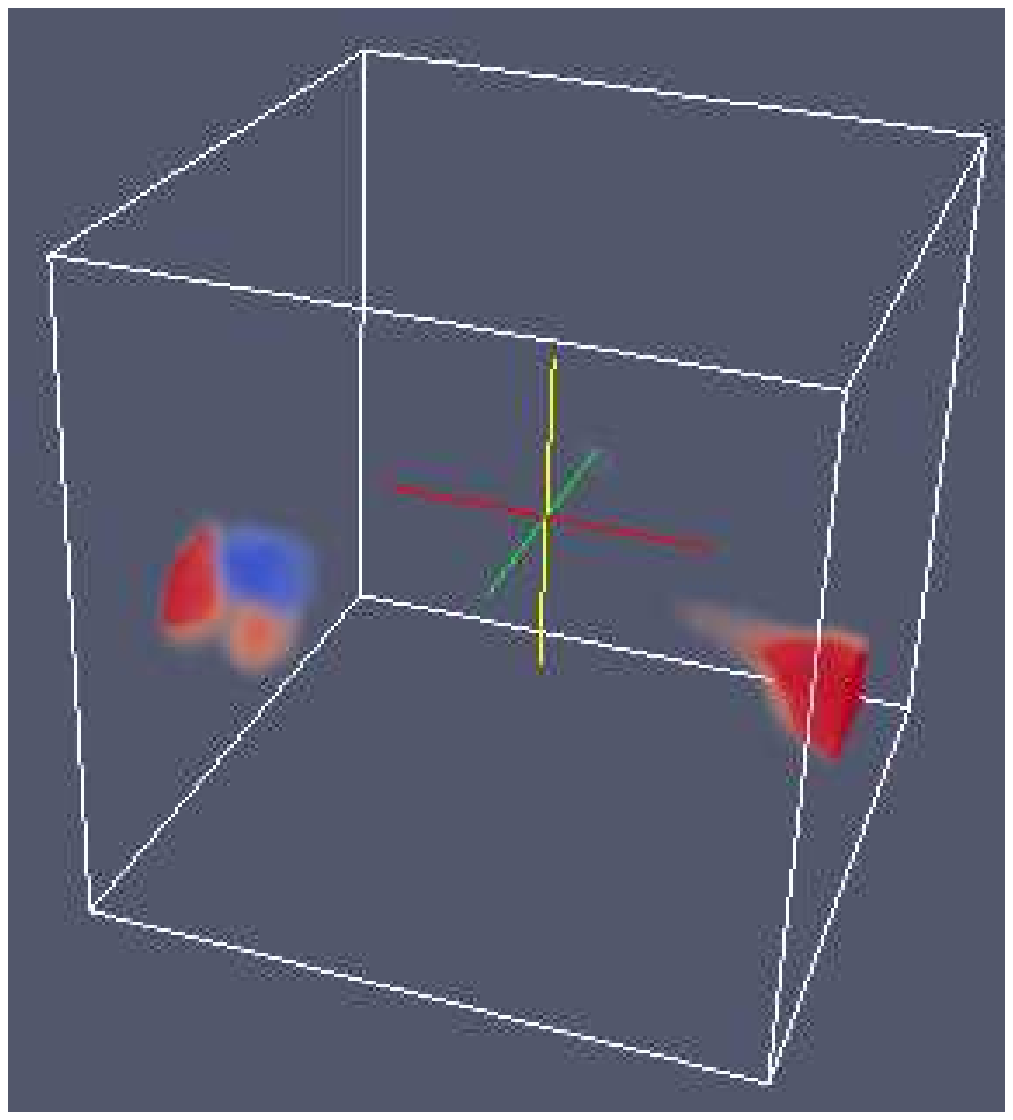}  \\
  \includegraphics[width=.18\textwidth]{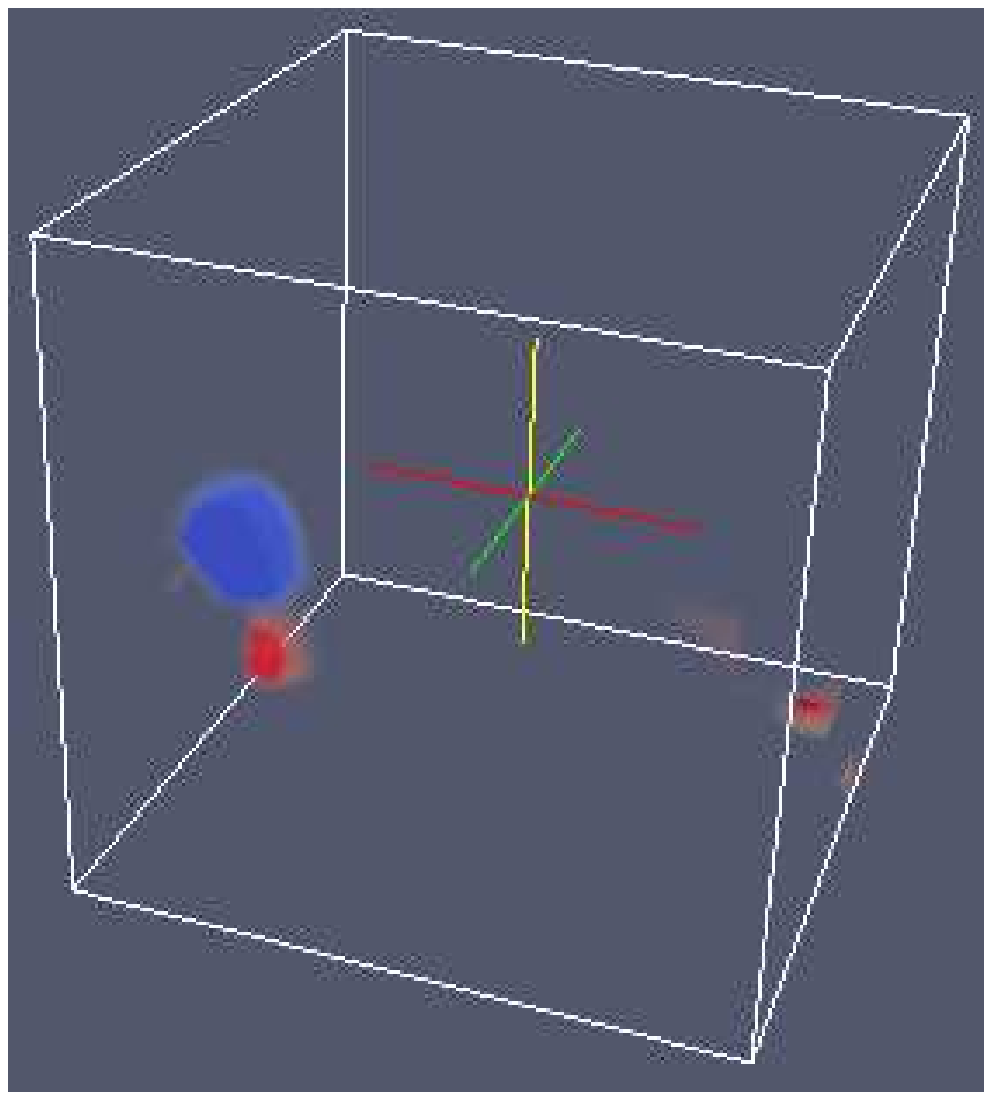}  &
  \includegraphics[width=.18\textwidth]{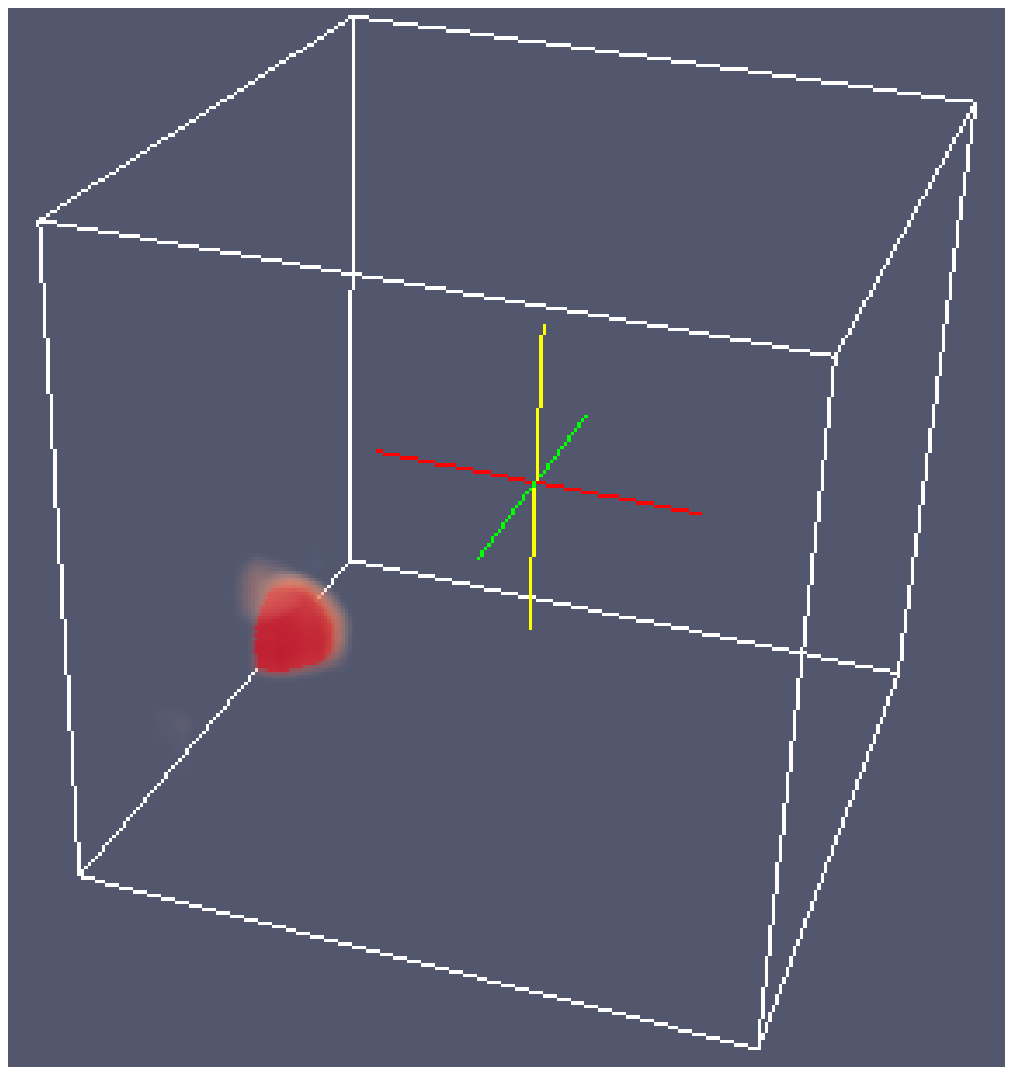}  &
  \includegraphics[width=.18\textwidth]{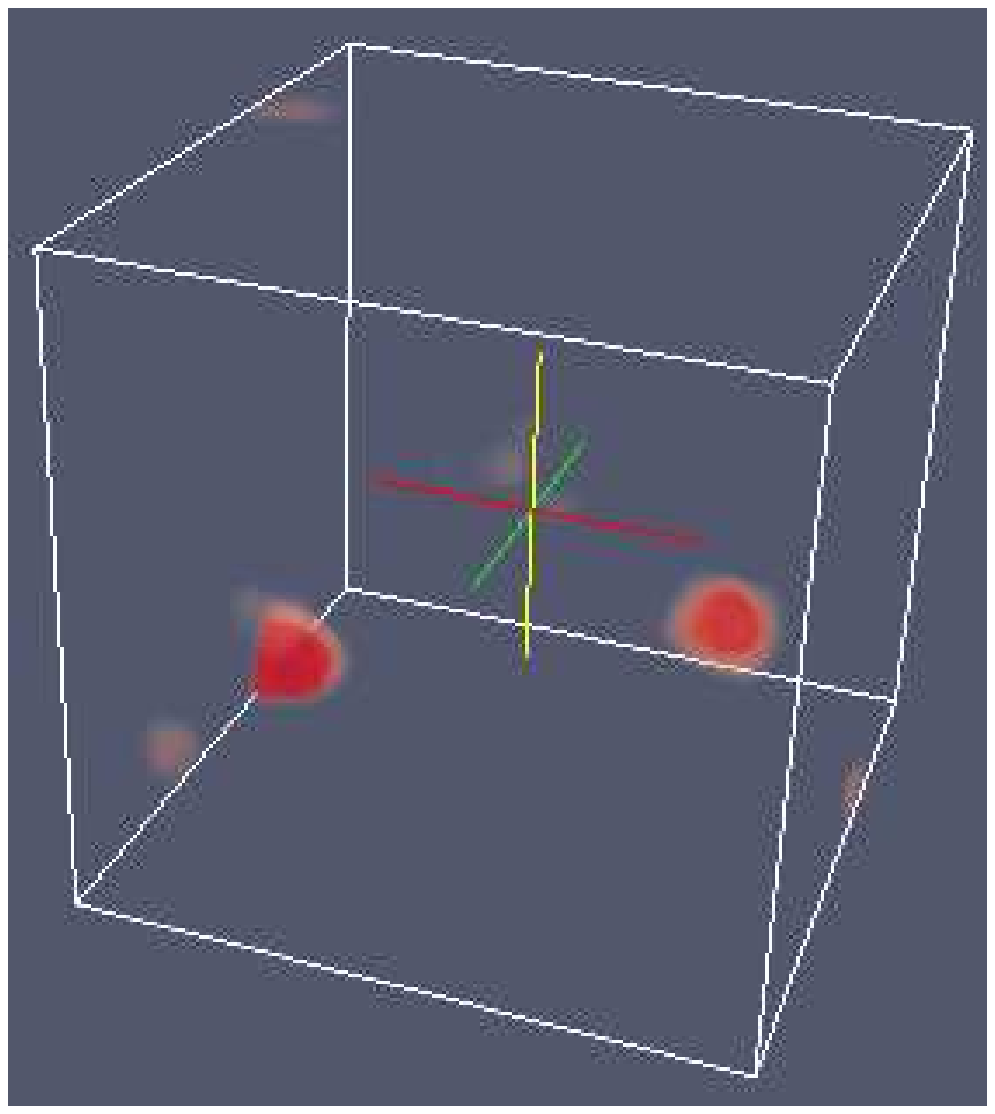}  &
  \includegraphics[width=.18\textwidth]{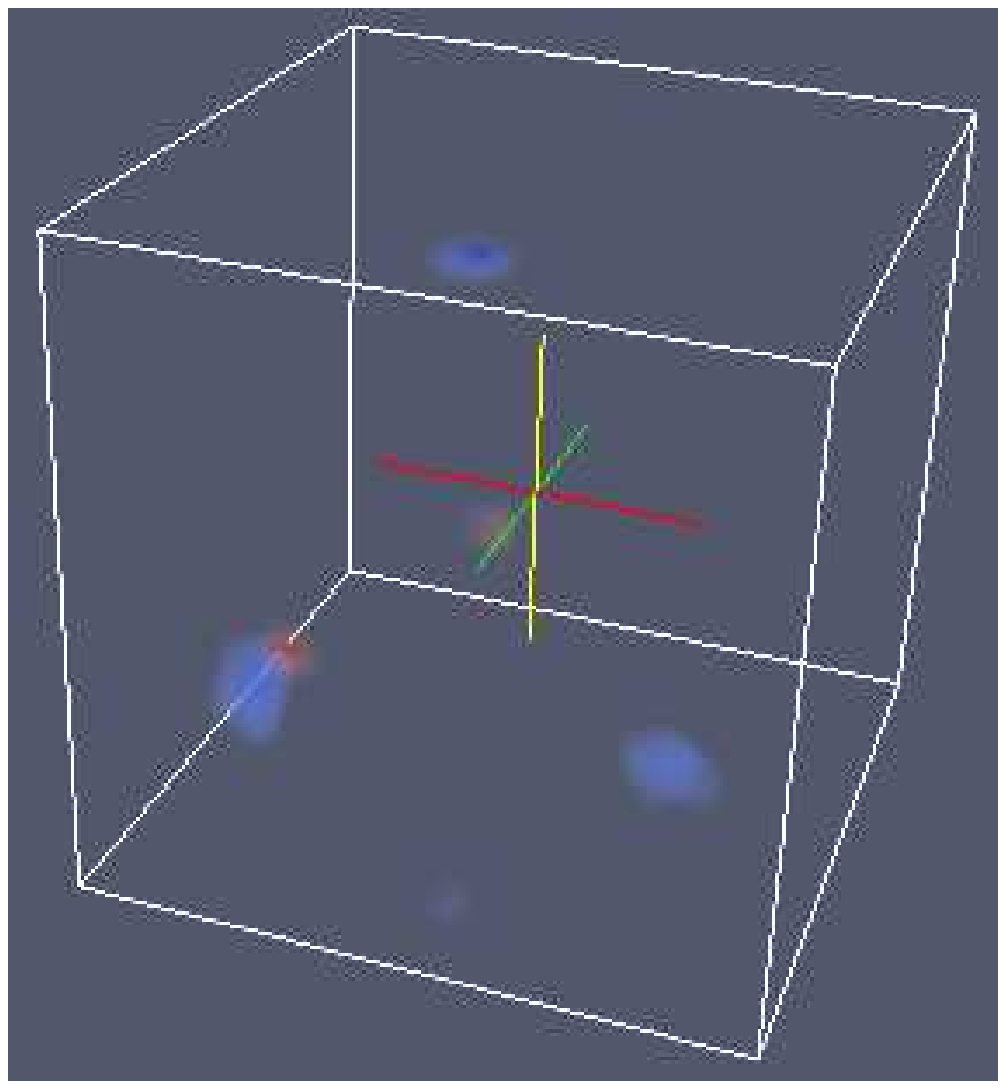}
\end{tabular}
\caption{(Euclidean) time evolution of the topological charge density 
of the two biggest clusters. \label{fig:1} }
\end{figure}

surprising that these
conditions, implemented by a Lagrange multiplier for each link,
do {\it not} prevent the lattice field to develop into a
smooth field free of UV fluctuations. Our numerical experiments have
shown that the ensembles obtained via constrained cooling {\it are}
smooth low-action configurations which can be easily studied by means of the
gluonic action density, the topological charge density and the modes of
%% EMI added "lowest" KL: removed ``lowest'', gluonic habe wir alle studiert
gluonic and fermionic operators.

\vskip 0.1cm
\textbf{Summary of main results:} It tuned out that these
configurations are suitable background fields for a semiclassical evaluation
of the partition function: an investigation of the gauge invariant
spectrum of the Yang-Mills Hessian~\cite{Huang:1996wj} shows a clear
separation between close-to-zero modes, which encode the collective
degrees of freedom of the semi-classical background, and the bulk of
gluonic modes, which
%% gives rise to its quantum weight (entropy).
merely counts towards its quantum weight (entropy).
%% EMI Frage: wuerde dieses quantum weight nicht das Gewicht der
%% EMI Konfiguration doppelt-zaehlen ? Der Gedanke mit der Entropie
%% EMI gefaellt mir aber.
%% KL  die kollektiven Moden sind in der Fluktuationsdeterminante
%%     enthalten; Aenderung siehe oben

\vskip 0.1cm
The topological charge of the streamline configurations can be determined
by simple means using the field-theoretic definition using the field-strenght
tensor. It turns out that the
topological susceptibility is protected by the constraints yielding
the value known for $SU(2)$ gauge theory~\cite{Teper:1985ek}.

\vskip 0.1cm
In order to trace out the spontaneous breakdown of chiral symmetry,
we calculated the near-zero eigenspectrum of the staggered quark operator
using backgrounds fields obtained by constrained cooling.
Although we have worked with staggered fermions for which the index theorem,
the close relationship between zero modes and topological charge is spoiled,
the presence of a band of close-to-zero modes proves that the chiral
symmetry is spontaneously broken. Here, the mechanism for spontaneous
symmetry breaking seems to be very much the same as for instanton
liquids: while a particular instanton gives rise to fermionic zero modes,
a liquid of instantons naturally induces a spectral density and hence
spontaneous symmetry breaking via the Banks-Casher relation.
We stress that, while the near-to-zero spectral properties of the {\it
streamline configurations}  are non-trivial, the spectral properties of
configurations emerging from standard cooling are largely different:
For the latter, the Creutz ratios probing confinement, the topological
susceptibility and the near-to-zero modes of the quark operator
are fading away with an increasing number of standard cooling iterations.

\vskip 0.1cm
%%
%% KL changed this paragraph
%%
The space-time texture of the topological charge gives the possibility
to confront the result of constrained cooling with any of the popular models.
A lattice field, originally prepared at $\beta=2.5$ with Wilson action,
underwent constrained cooling.
Table~\ref{tab1} lists its
$10$ biggest clusters sorted according to the modulus of the topological
charge $Q_\mathrm{lump}$ of the cluster. $N_\mathrm{lump}$ denotes the
number of space time points which belong to a particular cluster, and
$A_\mathrm{lump}$ denotes the action of the cluster. The total
topological charge of the configuration of $Q \approx 4$ is contributed
roughly from two large clusters with charge $15.7$ and $-11.2$.
Fig.~\ref{fig:1} 
shows the postive (red) and negative (blue) charge densities of just
these two clusters. All 24 time slices of a $24^4$ lattice are seen in
consecutive order. We refer
to~\cite{Langfeld:2010nm} for details on the cluster search.

By contrast
to a liquid of instantons, we found that individual clusters
{\it percolate}  and can have
any non-integer topological charge and that only the sum of all cluster
contributions is quantised (as it must be).
One can see how parts of the two clusters appear and disappear in the
successive time-slices.
%% EMI as a function of $x_4$.
%% EMI For this result, we have used the Wilson action with $\beta = 2.5$ and a
%% EMI simple plaquette based definition of the topological charge
%% EMI density (see~\cite{Langfeld:2010nm} for details).
%% EMI Schon gesagt
%% EMI Der Rest ist vorgezogen
%% For the configuration at hand, table~\ref{tab1} lists the
%% $10$ biggest clusters sorted according to the modulus of the topological
%% charge $Q_\mathrm{lump}$ of the cluster. $N_\mathrm{lump}$ denotes the
%% number of space time points which belong to a particular cluster, and
%% $A_\mathrm{lump}$ denotes the action of the cluster. Again, we refer
%% to~\cite{Langfeld:2010nm} for details. The total topological charge
%% of the configuration of $\approx 4$ roughly emerges from two large
%% clusters with charge $15.7$ and $-11.2$. By contrast to an ensemble
%% of instantons, we found that an individual cluster can have
%% any fractional topological charge and that only the sum of all
%% cluster contributions is quantised (as it must be).

\begin{table}
\begin{tabular}{c|ccc}
 num & $N_\mathrm{lump}$ & $Q_\mathrm{lump}$ & $A_\mathrm{lump}$  \\ \hline
             1 &     171726 &   0.1573E+02 &   0.3491E+02 \\
             2 &     158554 &  -0.1125E+02 &   0.2904E+02 \\
             3 &          7 &  -0.2017E-03 &   0.1720E-02 \\
             4 &          4 &   0.1919E-03 &   0.1177E-02 \\
             5 &          7 &   0.1777E-03 &   0.1315E-02 \\
             6 &         10 &  -0.1727E-03 &   0.1839E-02 \\
             7 &          3 &   0.1726E-03 &   0.8142E-03 \\
             8 &          6 &   0.1385E-03 &   0.1192E-02 \\
             9 &          4 &   0.1378E-03 &   0.7320E-03 \\
            10 &          5 &  -0.1354E-03 &   0.7577E-03
\end{tabular}
  \caption{Data of the visualised configuration. \label{tab1} }
\end{table}

%%%%%%%%%%%%%%%%%%%%%%%%%%%%%%%%%%%%%%%%%%%%%%%%
%% BACKMATTER
%%%%%%%%%%%%%%%%%%%%%%%%%%%%%%%%%%%%%%%%%%%%%%%%

%%
%% KL brauchen wir das hier?
%{\bf Acknowledgments:}



% \bibliographystyle{aipprocl} % if natbib is missing


%\bibliography{mybib2}{}
%\bibliographystyle{aipproc}

\end{document}